\documentclass[manuscript,screen]{acmart}
\AtBeginDocument{%
  \providecommand\BibTeX{{%
    \normalfont B\kern-0.5em{\scshape i\kern-0.25em b}\kern-0.8em\TeX}}}

\copyrightyear{2022}
\acmYear{2022}

\setcopyright{rightsretained}
\acmConference[HT '22]{Proceedings of the 33rd ACM Conference on
Hypertext and Social Media}{June 28-July 1, 2022}{Barcelona, Spain}
\acmBooktitle{Proceedings of the 33rd ACM Conference on Hypertext and
Social Media (HT '22), June 28-July 1, 2022, Barcelona,
Spain}
\acmDOI{10.1145/3511095.3536361}
\acmISBN{978-1-4503-9233-4/22/06}

\usepackage{xspace}
\usepackage{stackengine,graphicx} %
\setstackgap{S}{-1.5pt}

\newcommand{\shorttitle}{Short title}
\newcommand{\projectname}{{Open Source Attention}\xspace}
\newcommand{\shortprojectname}{{OSA}\xspace}

\newcommand{\xhdr}[1]{\vspace{1mm}\noindent{{\bf #1.}}} %

\begin{document}

\title[From Users to (Sense)Makers]{From Users to (Sense)Makers: On the Pivotal Role of Stigmergic Social Annotation in the Quest for Collective Sensemaking}

\author{Ronen Tamari}
\email{ronent@cs.huji.ac.il}
\orcid{0000-0002-6049-591X}
\affiliation{%
  \institution{DAOStack, Hebrew University of Jerusalem}
  \country{Israel}
}

\author{Daniel A Friedman}
\email{danielarifriedman@gmail.com}
\affiliation{%
  \institution{University of California, Davis}
  \country{USA}
}

\author{William Fischer}
\email{will@veeo.io}
\author{Lauren Hebert}
\email{lauren@veeo.io}
\affiliation{%
  \institution{Veeo}
  \country{USA}
}

\author{Dafna Shahaf}
\email{dshahaf@cs.huji.ac.il}
\affiliation{%
  \institution{Hebrew University of Jerusalem}
  \country{Israel}
}

\renewcommand{\shortauthors}{Tamari, Friedman, Fischer, Hebert and Shahaf}

\begin{abstract}

The web has become a dominant epistemic environment, influencing people’s beliefs at a global scale. However, online epistemic environments are increasingly polluted, impairing societies’ ability to coordinate effectively in the face of global crises. We argue that centralized platforms are a main source of epistemic pollution, and that healthier environments require redesigning how we collectively govern attention. Inspired by decentralization and open source software movements, we propose Open Source Attention, a socio-technical framework for ``freeing'' human attention from control by platforms, through a decentralized eco-system for creating, storing and querying stigmergic markers; the digital traces of human attention.
\end{abstract}


\maketitle

\section{Introduction}

The web has become a dominant epistemic environment, shaping peoples’ beliefs and knowledge on a global scale. The web, however, is also currently a severely polluted epistemic environment~\cite{levy2021bad}, due to highly centralized and opaque information ecologies, coupled with incentive misalignment and unprecedented information overload. A small number of major web platforms such as Google and Facebook have gained immense control over the means to search, create, and distribute information~\cite{lorenz2020behavioural}. Centralization leads to opacity, in which network data as well as algorithms for content creation, search, and distribution are effectively hidden away from public, scientific, and ethical oversight~\cite{bakcoleman2021}.

Platform incentives are fundamentally misaligned with those necessary for healthier epistemic environments~\cite{zuboff2019age}. For example, centralization and control of data are necessary for running lucrative ``attention markets'', but ultimately hinder attempts to address information overload, and undermine both user autonomy~\cite{Shah2022Situating} as well as the open information networks necessary for healthy democracies~\cite{Zuckerman2020,Kozyreva2020}. Platforms are implicated in a host of problematic social phenomena, including the spread of false information, behavioral changes, and societal polarization, epistemic distraction and degradation of individual and collective sense-making capacities~\cite{williams2018stand,lorenz2020behavioural}. Impending global ecological and societal crises lend increased urgency to addressing these problems: astute collective sense- and decision-making have perhaps never been more needed~\cite{williams2018stand,West2021,bakcoleman2021}.

Laudable recent efforts have called attention to this precarious state of affairs of collective online sensemaking ~\cite{lorenz2020behavioural,Kozyreva2020,bakcoleman2021}. However, while providing invaluable insights, they have largely focused on improving platforms through regulatory action, whether internally or externally imposed. Such platform-centric approaches are an important step towards healthier epistemic environments, but face inherent limitations (\S\ref{sec:opensourcestig}), and are fraught with many impediments as they often run counter to powerful platforms’ core business models. Perhaps more crucially, platform-centric initiatives cannot adequately account for the fundamentally distributed~\cite{Gureckis2006ThinkingIG}, self-organizing~\cite{Franks1997SelforganizingNC}, and stigmergic~\cite{Marsh2008} nature of collective intelligence.

We argue that reflecting these considerations in practice would benefit from a more radical redesign of our epistemic environments, centered around guiding principles of agency, transparency, interoperability, decentralization, and a collective conceptual transition to a ``maker'' mindset~\cite{KOSTAKIS2015126}; from passive users, to more active (sense)makers.

Inspired by both theoretical and practical breakthroughs of decentralization and open source software movements contra entrenched centralized systems, we propose \projectname (\shortprojectname), a conceptual framework and ``call to movement'' towards decentralized, open-source, stigmergic annotation. We envision our framework as a step towards systems for distributed governance, education, and control of collective sense-making and attention.

Hypertext and social annotation play a pivotal role in our proposed transition: in the current \emph{platform-centric} ecology, user annotations (such as likes, retweets, etc) are locked across platforms’ data siloes where they serve to optimize \emph{platform} growth. \shortprojectname aims to empower \emph{maker-centric} ecologies by employing distributed content creation and storage technology (e.g., Solid~\cite{solid2016}). In this way, makers will control creation and dissemination of their annotations, which can then be leveraged to optimize personalized \emph{human} growth and learning for individuals and collectives.

\begin{figure*}[b]
  \centering
  \includegraphics[width=0.84\linewidth,scale=0.50]{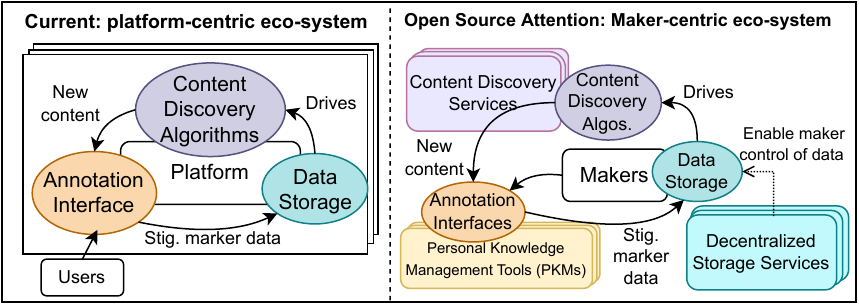}
  \caption{Platforms leverage control of both data and content discovery algorithms to drive growth at the expense of users (left); Decoupling data and algorithms incentivizes content discovery services oriented towards human-centered growth (right). }
  \Description{Overview of Open Source Attention eco-system }
  \label{fig:osa}
\end{figure*}

\section{Distributed, stigmergic foundations of collective sense-making}
Sense-making refers to processes by which agents make sense of their environment, achieved by organizing sense data until the environment is understood well enough to enable reasonable decisions~\cite{weick1995sensemaking}. 
Theories of extended~\cite{clark1998extended} and stigmergic~\cite{Marsh2008} cognition highlight the integral role of \emph{environment modification} in sense making; agents actively change their environment to assist internal cognitive processes (e.g., writing to-do notes) as well as indirect \emph{stigmergic} communication with others (e.g., ant pheromone trails). Stigmergy is particularly relevant for the setting of collaboration of large-scale groups~\cite{Elliott_2006}. In stigmergic communication, the environment acts as a kind of distributed memory; modifications left by others provide cybernetic feedback, driving both \emph{emergence} of novel system-level behavior from local interactions of agents, and \emph{immergence} (individual interactions informed by a global state of affairs)~\cite{Marsh2008}. Sense-making is thus inherently co-created, through agents modifying their environment and reacting to changes made by others.

What kind of environment modifications are relevant to consider for sense-making in vast digital spaces? The literature broadly distinguishes between two types of modifications: \emph{sematectonic stigmergy}, which directly alters the environment state (in the digital case: creating new content, such as publishing a blog post), and \emph{stigmergic markers}, which do not directly modify content, but rather serve as signalling cues (in the digital case: likes, annotations, hyperlinking of text). Importantly,  stigmergic markers play a central role in assessing epistemic quality of content, both for humans~\cite{Marsh2008,levy2021bad} and machines~\cite{lorenz2020behavioural}, due to the sheer volume of information as well as challenges in endogenous content interpretation. Stigmergic markers may be explicitly left by users (e.g., likes) or implicitly recorded through their behavior (e.g., link click-through data, reading time).

\section{Open-sourcing stigmergic markers for healthier epistemic environments}
\label{sec:opensourcestig}

Polluted epistemic environments are often framed as casualties of the ``attention economy''; platforms selling user data to advertisers and putting up ads in social media feeds, with the aim of ``capturing'' users’ attention and seducing them to make yet another purchase. While ``data'' and ``attention'' are popular abstractions, the stigmergic perspective is valuable in guiding practical redesign of epistemic environments. Stigmergic markers can be thought of as digital traces of human attention, whose primacy as indicators of epistemic value makes them precious resources, whether for extractive (e.g., ad-tech) or constructive (e.g., collective sense-making) purposes. In the following sections, we illuminate the role of stigmergic markers in nourishing healthier epistemic environments.

\subsection{From attention to intention}
Healthier epistemic environments involve moving from exploitation of \emph{a}ttention to supporting our \emph{in}tentions~\cite{williams2018stand}. This transition requires two paradigm shifts. First, a mindset shift on the human side, from passive, unwitting users consuming ``unhealthy information diets''~\cite{johnson2011information}, to active makers, who cultivate growth-oriented intentions and are mindful of the (stigmergic) traces they leave, as well as their role as co-creators in the larger digital and physical ecology. Realistically, humans stand no chance of making the transition in isolation; content discovery algorithms are indispensable for navigating vast digital landscapes, but to a large degree are controlled by platforms~\cite{lorenz2020behavioural}. Accordingly, the second shift involves re-designing our epistemic environments to support this transition by empowering makers through human-centric content discovery. As shown in Fig. \ref{fig:osa}, current content discovery is platform-centric: platforms enjoy a closed feedback loop consisting of both the content discovery algorithms as well as the stigmergic marker data needed to drive algorithmic optimization towards platform growth~\cite{verborgh_timbl_chapter_2022}. Human-centric content discovery requires supplanting this degenerative cycle with a more symbiotic information ecology, in which makers create and control their stigmergic markers, and thus are empowered to share their data to content discovery services oriented towards personalized individual or collective growth. Content moderation is an important representative example~\cite{Masnick2019}: moderation is intractable in centralized systems, due to inherent limitations of AI capabilities as well as the scale of complex human adjudications needed. In contrast, decentralized eco-systems enable a ``marketplace of filters'', where different individuals and organizations can create and tune content moderation systems for their own needs.

\subsection{Open Source Attention: maker-centered information ecology}

Analogously to open-source code and common domain knowledge~\cite{KOSTAKIS2015126}, stigmergic markers can be thought of as a public good. However, despite their unique importance, surprisingly little work has specifically targeted their decentralization (\S\ref{sec:discussion}). Stated simply; where open source software is a movement to ``free'' software, similarly \shortprojectname is a movement to ``free'' stigmergic markers, starting from basic hypertext primitives: emotional valence (e.g., likes), bi-directional links, span highlighting, semantic categorization (tags, bookmarks), and textual annotation. We envision decentralized, maker-centered ecologies, comprised of three main architectural elements (see also Fig. \ref{fig:osa}):

\xhdr{Annotation tools} Enable makers to easily create markers attached to any URL or content element included therein, not just where platforms provide like buttons~\cite{capadisli2017decentralised}. Some types of markers should themselves be mark-able, allowing for example the option to ``like'' a particular annotation, or link between two annotations. Future extensions can address implicit stigmergic markers such as read-time or click-through counts~\cite{Marques2013}. Apps recording these function as automatic annotation tools, though their implicit nature requires extra caution with regard to consent and data privacy issues. 

\xhdr{Self-sovereign storage} Makers own their markers and control their visibility (private, public, etc) to other people or services. Identity provision is a key related service that can (but does not have to be) provided along with storage~\cite{verborgh_timbl_chapter_2022}.

\xhdr{Content discovery services} Rather than platforms' monolithic and opaque feeds, a decentralized ecology encourages a market of diverse and human-centered content discovery services. For example, competing interfaces for social media that better moderate trolls, promote thought-provoking stories, or provide customizable feed controls~\cite{Masnick2019}.

\section{Discussion}
\label{sec:discussion}

While the idea of leveraging stigmergic markers for collective sense-making has a long history~\cite{bush1945we}, most contemporary open-source and decentralization efforts have focused on sematectonic (content-creating) stimergy, such as code, social media, financial ledgers and executable contracts~\cite{CASINO201955}. Closest to our proposal is the Solid eco-system~\cite{solid2016} that, similarly, targets ``re-decentralizing the web''~\cite{verborgh_timbl_chapter_2022}, and empowers individuals to control their data. Solid also features a marketplace of services, including the dokieli decentralized annotation client for scientific research~\cite{capadisli2017decentralised, capadisli2020linkedphd}. While Solid and dokieli are inspiring initial steps, they are limited with regard to content discovery services or social incentives. More broadly, where Solid is primarily a technology, \shortprojectname proposes an ecological perspective accounting for the embeddedness of such technologies in wider social, educational and economic contexts. For  example, a key extra-technological challenge concerns changing norms around knowledge work. Similarly to how platforms changed the culture around certain kinds of content creation, effectively turning us all into performers, well designed social networks could help shape the norms and prestige associated with sense-making activities. Academic Twitter demonstrates that even without direct economic incentives, social incentives lead experts to freely share high-quality information publicly~\cite{actwit2020}.

Another key question concerns scale: for global-scale sense-making, any proposal must necessarily compete with massive, well established platforms. While recent years are seeing a resurgence of \emph{personal} knowledge management apps (PKMs) enabling content creation and annotation, knowledge tends to remain siloed at the individual level; adaptation to \emph{collective} knowledge management (CKM) has been limited.\footnote{\url{https://athensresearch.ghost.io/season-2/}} \shortprojectname is naturally congruent with the promising ``protocols, not platforms'' approach~\cite{Masnick2019}; rather than head-to-head competition between PKMs, existing PKM growth can be bootstrapped for CKM by introducing interoperable protocols and storage for stigmergic primitives (e.g., links, tags). In this way, data from across diverse PKM apps could be shared to contribute to collective sense-making efforts.

Many open questions remain out of scope of this short piece, which is best seen as a call to attention; success in surmounting the formidable challenges faced today by humanity requires that ``we give the right sort of attention to the right sort of things''~\cite{williams2018stand}. We have claimed that attending ``to the right things'' will require re-imagining our socio-technological systems for governing collective attention; we hope our proposal will help galvanize action towards this vital cause.

\begin{acks}
We thank Zak Stein for inspiring our exploration, and we thank Nimrod Talmon and the DAOStack team for thoughtful feedback and support. We also thank Metagov and RadicalXChange for cultivating the wonderful real-world and online spaces that seeded this collaboration.
\end{acks}

\bibliographystyle{ACM-Reference-Format}
\bibliography{sample-base}

\end{document}